\documentclass[aps,prb,twocolumn,showpacs,superscriptaddress,preprintnumbers,amsmath,amssymb,longbibliography]{revtex4-1}

\usepackage{gensymb}                               % Symbol degree : \degree
\usepackage{xcolor}
\usepackage{ulem}
\usepackage{graphicx}
\usepackage{array,multirow,makecell}
\usepackage{mhchem}
\usepackage{bm}
\usepackage{tikz}
\usepackage[pdftex,colorlinks=true,pdfstartview=FitV,linkcolor=blue,citecolor=blue,urlcolor=blue]{hyperref}

\begin{document}

\title{Colloidal Synthesis of Nanoparticles: from Bimetallic to High Entropy Alloys}

% Author: Please give full first and last names for authors and include * after the name of all corresponding authors
\author{Cora Moreira Da Silva}
\affiliation{Laboratoire d'Etude des Microstructures, ONERA-CNRS, UMR104, Universit\'e Paris-Saclay, BP 72, Ch\^atillon Cedex, 92322, France}

\author{Hakim Amara}
\affiliation{Laboratoire d'Etude des Microstructures, ONERA-CNRS, UMR104, Universit\'e Paris-Saclay, BP 72, Ch\^atillon Cedex, 92322, France}
\affiliation{Universit\'e de Paris, Laboratoire Mat\'eriaux et Ph\'enom\`enes Quantiques (MPQ), F-75013, Paris, France}

\author{Fr\'ed\'eric Fossard}
\affiliation{Laboratoire d'Etude des Microstructures, ONERA-CNRS, UMR104, Universit\'e Paris-Saclay, BP 72, Ch\^atillon Cedex, 92322, France}

\author{Armelle Girard}
\affiliation{Laboratoire d'Etude des Microstructures, ONERA-CNRS, UMR104, Universit\'e Paris-Saclay, BP 72, Ch\^atillon Cedex, 92322, France}
\affiliation{Universit\'e Versailles Saint-Quentin, U.~Paris-Saclay, Versailles, 78035, France}

\author{Annick Loiseau}
\affiliation{Laboratoire d'Etude des Microstructures, ONERA-CNRS, UMR104, Universit\'e Paris-Saclay, BP 72, Ch\^atillon Cedex, 92322, France}

\author{Vincent Huc}
\affiliation{Institut de Chimie Mol\'eculaire et des Mat\'eriaux d'Orsay, CNRS, Paris Sud, U.~Paris-Saclay, Orsay, 91045, France}

\begin{abstract}

At the nanometric scale, the synthesis of a random alloy (\textit{i.e.} without phase segregation, whatever the composition) by chemical synthesis remains a not easy task, even for simple binary type systems. In this context,  a unique approach based on the colloidal route is proposed enabling the synthesis of face-centred cubic and monodisperse bimetallic, trimetallic, tetrametallic and pentametallic nanoparticles with diameters around 5~nm as solid solutions. The Fe-Co-Ni-Pt-Ru alloy and its subsets are considered which is a challenging task as each element has fairly different physico-chemical properties. Nanoparticles are prepared by temperature-assisted co-reduction of metal acetylacetonate precursors in the presence of surfactants. It is highlighted how the correlation between precursors' degradation temperatures and reduction potentials values of the metal cations is the driving force to achieve a homogenous distribution of all elements within the nanoparticles.

\end{abstract}

\maketitle 

% Text: Please use section headings and subheadings as specified below. For communications, all section headings apart from Experimental Section should be removed
% Please make the first reference to a display item bold: \textbf{Figure 1}
% Do not abbreviate Figure, Equation, etc.; display items are always singular, i.e., Figure 1 and 2.
% Equations are always singular, i.e., Equation 1 and 2, and should be inserted using the {equation} environment, not as graphics
% Please do not use footnotes in the text, additional information can be added to the Reference list.

\section{Introduction}

Nowadays, metallic nanoparticles (NPs) are widely studied in many combined scientific fields for their specific reactivity and/or physical properties resulting from their reduced size. Their nanometric dimension gives them different properties compared with their bulk counterparts~\cite{Alloyeau2012}. This is particularly true in case of bimetallic alloys (or nano-alloys) where the combination of two metals within a nanoparticle could lead to very high synergistic effects, attractive in many fields such as catalysis reactions, optics, optoelectronics and medical applications~\cite{Ferrando2008}. A particularly exciting development is the extension to nanoalloys with more than two metals present at the same time to form multicomponent metallic alloys usually called high-entropy alloy (HEA) nanoparticles. With their dynamic expansion, high entropy alloys (HEAs) in bulk form~\cite{Cantor2004, Yeh2004}, have become an active area of research in the materials science community, particularly for their mechanical properties~\cite{George2020}. These materials, containing multiple elements (4 to 6) in a composition range of 5-35 at.\%~\cite{Miracle2017}, tend to form simple solid solution microstructures partly due to their configurational entropy that increases with the number of elements. This class of random alloys with impressive strength and toughness~\cite{Varvenne2016} is foreseen as a new design axis in metallurgy to control a wide range of mechanisms~\cite{Nohring2020}. Four fundamental effects are put forward to explain the extraordinary properties of bulk HEA: high configurational entropy, large lattice distortion, slow diffusion of atoms, and cocktail effects~\cite{Miracle2017}. Due to the remarkable changes demonstrated by pure metals and alloys downsizing  in a wide range of fields, the concept of HEA has naturally extended to nanoscale. Although research on HEA NPs is still in its infancy, very promising results have revealed their unique physical and chemical properties, expecting to exceed those of conventional metallic nanoalloys in various applications. Typical example includes quinary PtPdRhRuCe NPs as ammonia oxidation catalysts~\cite{Yao2018} or oxidized FeCoNiCuPt HEA for applications targeting H$_{2}$ formation and storage~\cite{Song2021}. As a result, this spectrum of impressive properties makes this new broad class of structural materials attractive~\cite{Loffler2019, Lacey2019, Xie2019, Gao2020, Li2020b, Fu2021, Feng2021, Bartenbach2021}. To take full advantage of their potential as structural and functional materials for future nanotechnologies, it is crucial to thoroughly understand the structure-property relationship in HEA-NPs. Consequently, the development of innovative synthesis protocols is required to gain control over the size, shape composition and structural phase of HEA NPs. In this context, high temperature (HT) methods such as carbothermal shock and plasma arc discharge syntheses have been proposed with varying degrees of efficiency~\cite{Yao2018, Waag2019}. Despite these outstanding successes, the very extreme synthesis conditions make it difficult to control the resulting HEA NPs, which is an obstacle to scaling up their use in various industrial applications. Besides, more appropriate approaches to produce HEA NPs in soft conditions have been proposed deriving from wet-chemistry synthesis~\cite{Huang2015, Singh2015, Li2019}. However, most studies have resulted in NPs with poor control of size~\cite{Wu2020}, monodispersity~\cite{Bondesgaard2019}, shape~\cite{Li2020b} or require a post-processing steps~\cite{Chen2021}. Recently, several ternary systems in solid solution have been synthesized using colloidal route to highlight their potential in various catalytic applications~\cite{Bondesgaard2019, Li2019, Li2020b}. If more elements are present, structures with undesirable phase separation, are obtained~\cite{Chen2016}. Despite these relative achievements~\cite{Zhang2021, Li2020b, Li2019b}, little effort has been devoted to the synthesis of NPs and more precisely to the relevance of the experimental parameters impacting the physico-chemical properties of the obtained NPs.\\

In the present work, we present a simple and versatile one-pot approach to produce HEA NPs under soft synthesis conditions ensuring the formation of simple solid solution microstructures, the structural signature of HEAs. To achieve this goal, we extend our new colloidal procedure for the synthesis of bimetallic Ni$_{x}$Pt$_{1-x}$ with controlled size and chemical composition to fabricate HEA NPs~\cite{MoreiraDaSilva2020}. A major difficulty in optimising the synthesis parameters is to characterise the obtained NPs. Typically, X-Ray Diffraction (XRD) analyses can been used but in case of small NPs, very broad and low intensity peaks are observed~\cite{Li2019} making the analysis difficult for elements with the same crystallographic structure and close lattice parameters as discussed in the present work. Here, the structural properties of the synthesized NPs are studied by High-Resolution transmission Electron Microscopy (HRTEM) using different modes. More precisely, to assert that the NP do not present any phase segregation and to quantify their chemical compositions, Energy Dispersive X-Ray analysis (EDX) are performed at the single particle level. The targeted material is the Fe-Co-Ni-Pt-Ru alloy (and its subsets), a perfect model system having several advantages to highlight the potential of our method. Indeed, this particular HEA alloy is based on individual elements crystallizing into various structures such as hcp, fcc and bcc. Moreover, it contains elements across different transition-metal series (3$d$, 4$d$ and 5$d$) with a broad disparity of atomic radius. All these factors make complex and challenging the elaboration of Fe-Co-Ni-Pt-Ru NPs in a FCC solid solution with controlled size and shape. 

\section{Colloidal synthesis of HEA NPs}

The colloidal route is widely used in case of pure NPs as an efficient and accurate way to obtain nano-objects with controlled structure (size and shape)~\cite{Murray2001a,Tanh2014,Kobayashi2015}. It is based on the reduction of metallic precursors by a reducing agent and the presence of surfactants immersed in an organic environment at a given temperature~\cite{Zhang2019}. Using this colloidal route, bimetallic nickel-platinum NPs have been recently obtained with a specific controlled chemical composition, ranging from pure Ni to pure Pt and well-defined size below 5 nm~\cite{MoreiraDaSilva2020}. More precisely, the synthesis temperature proved to be crucial to produce NPs in solid solution phase, \textit{i.e.} without phase segregation such as core-shell or Janus structures. After such a success, we will seek to extend this synthesis route to overcome an additional difficulties and achieve multi-elements (from 3 to 5 in solid solution), truly allowed NPs. \\

\begin{figure}[h!]
\begin{center}
\includegraphics[width=1.00\linewidth]{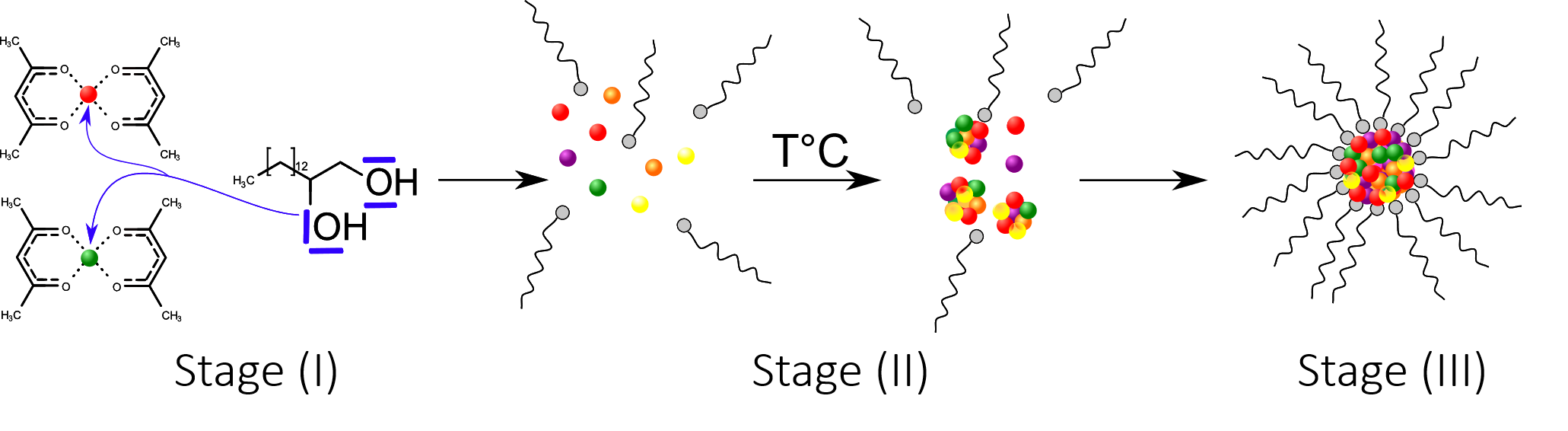}
 \caption{Mechanism of colloidal synthesis of metal nanoparticles. Stage (I) : metallic cations reduction. Stage (II) : nucleation of thermally activated NPs. Stage (III) : Stabilisation of NPs by surfactant binding.}
\label{mechanism}
\end{center}
\end{figure}
As seen in Figure~\ref{mechanism}, colloidal synthesis involves three key stages driven by experimental parameters (redox potential, temperature and surfactants). The first one corresponds to the reduction of metallic precursors containing metal cations ($M^{n+}$) down to their zero oxidation states by a reducing agent (here a diol) as electrons source. Such mechanism is rationalised by the redox potential ($E^{0}$) of the different species involved, a helpful empirical value to predict the tendency of chemical species to acquire or lose electron(s). %For a given redox couple, $E^{0}$ is determined in a standard way using a set-up of two half-cells connected by a salt bridge at atmospheric pressure and room temperature. 
In the case of the bimetallic NPs synthesis, the redox reaction involves three redox couples (two metal cations and the reducing agent) whose classification by their potential $E^{0}$ enables to predict which reaction is possible. More precisely, such reaction can only be achieved between the strongest oxidant of a couple with the strongest reductant of the other couple. This is perfectly illustrated by the gamma rule which can forecast the direction of the reaction as well as its relative kinetics.  
\begin{figure}[!h] \centering
\begin{center}
\includegraphics[width=1.00\linewidth]{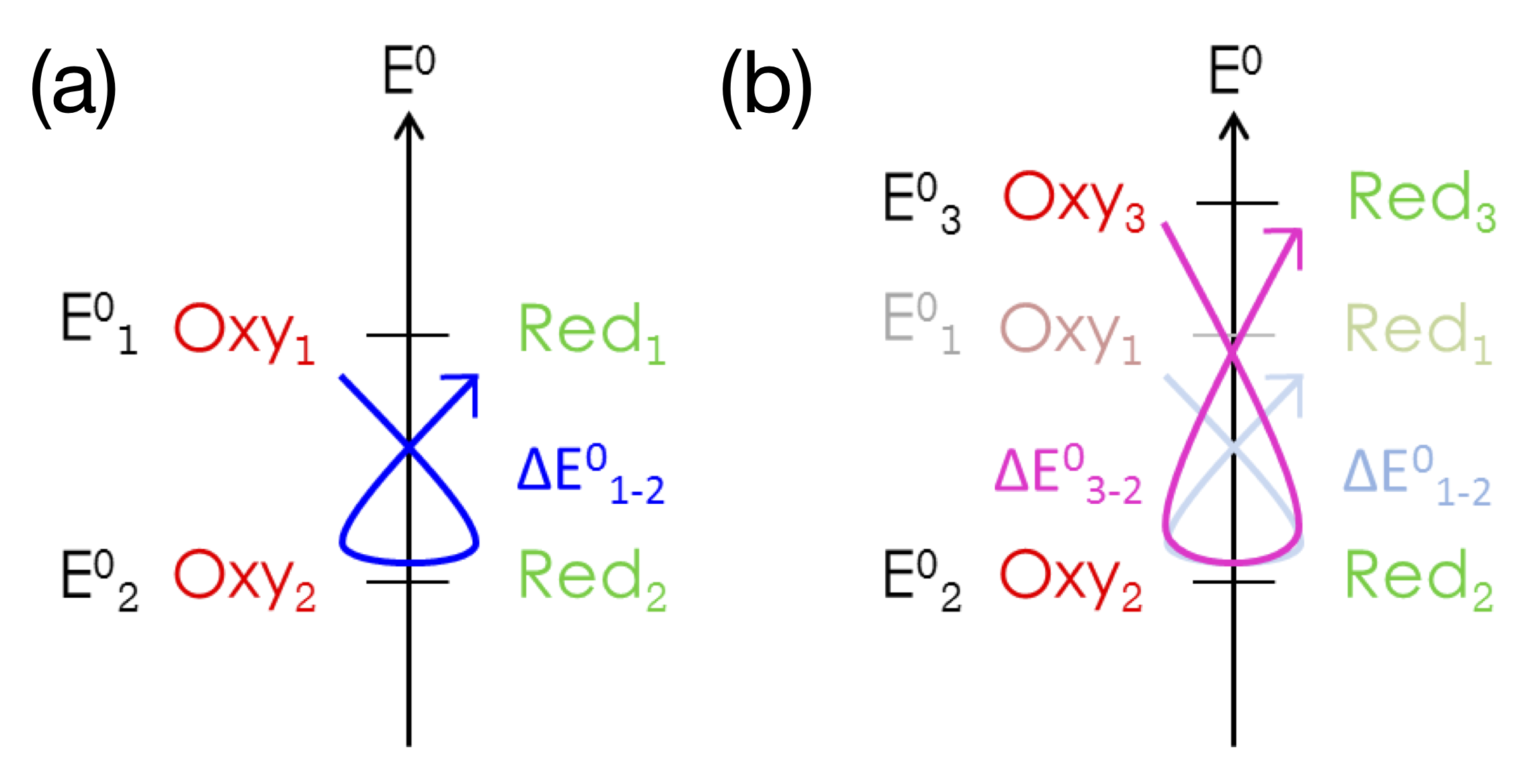}
\caption{Direction prediction, probability and kinetics of a redox reaction using the \textit{gamma} rule for a reaction with (\textbf{a}) two and (\textbf{b}) three redox couples.}
\label{gamma}
\end{center}
\end{figure}
First, let us consider two redox couples in a solution between a metallic cation ($Oxy_1/Red_1$) and the reducing agent ($Oxy_2/Red_2$) with standard redox potentials $E^0_1$ and $E^0_2$, respectively such as $E^0_1 > E^0_2$ and $\Delta E ^{0}_{1-2}= E^0_2-E^0_1$. Following the gamma rule (see Figure~\ref{gamma}a), the stronger oxidant ($Oxy_1$) will react with the stronger reductant ($Red_2$) to produce $Oxy_2$ and $Red_1$ species. In this reduction process, the more positive $\Delta E ^{0}_{1-2}$ is, the more thermodynamically favored the following reaction is : $\ce{Oxy1 + Red2 -> Oxy2 + Red1}$. The reduction ability is also reflected in the degradation temperature of the precursor containing the cation. The more easily a cation is reduced, the lower its degradation temperature. Interestingly, the gamma rule can be extended to several redox couples in solution. As seen in Figure~\ref{gamma}b, we can predict that the redox reaction \ce{Oxy3 + Red2 -> Oxy2 + Red3} is the more thermodynamically and kinetically favourable compared to the others. At this point, note that the presence of ligands bonded to metal cations may influence this reduction reaction. According to Marcus' theory~\cite{marcus1964}, the redox potential of $M^{n+}/M^0$ pair is modified by the nature of the coordination sphere around the metal cations. This potential modification is function of the nature of the ligands, \textit{i.e.} the bond strength  between ligands and metal atoms. In case of strong bonds, a precursor stabilization, associated with a decrease of the redox potential of the cation is observed due to electrons density transfer from the ligand to the metal. To take these ligand effects into account, the metal precursors used have all the same coordination sphere made of acetylacetonate groups. Following this approach, different bimetallic NPs have been synthesized in the literature. Typical examples include AuRu~\cite{Zhang2018} and NiRu~\cite{Chen2012} NPs. This gamma rule can therefore be extended to several elements but very surprisingly such an approach has never been adapted to synthesize nanoparticles containing more than two metallic elements. At the end of the reduction process, a second phase starts, where the different zerovalent metal atoms previously produced bind together as small clusters that keep growing (as long as metallic atoms are still produced in solution) during the third phase of the synthetic process. This growth process gives rise to particles which turn out to be very unstable, due to their strong tendency to form aggregates precipitating out of the solution. To overcome this difficulty, the second and third steps of the colloidal synthesis make use of surfactants, able to bind to the surface of the metallic nanoparticules to stabilize them and make these nanoobjects soluble in various solvents (Figure~\ref{mechanism}). Under these conditions, NPs with controlled size and shape are obtained, with two key parameters : the nature of the surfactants and the temperature. Surfactants are amphiphilic species that link to specific facets to prevent their growth, thus controlling the morphology of the NPs~\cite{Peng2000, Peng2009, Ahmadi1996, Bratlie2007, Ren2007, Song2005, Qu2006}. It is also interesting to notice that the presence of surfactants prevents the coalescence of NPs allowing a better control of the size of the synthesized nanomaterials~\cite{Samia2006,Abbott2015}. Consequently, the choice of the surfactant is crucial in the colloidal synthesis. Regarding the role of the temperature in chemical synthesis in general, it is obviously considerable on NPs formation as it is involved in the nucleation-growth mechanism, ripening phenomena, solubilisation, etc.~\cite{Tanh2014}. \\

The synthesis of NPs with controlled structures (size, composition) is therefore a subtle balance involving many steps driven by experimental parameters that are necessarily interconnected, making the task even more delicate. A specific difficulty of our study is to obtain a solid solution of NPs, the typical signature of a HEA system. Recently, a quinary NP alloy consisting of five platinum group
metals have been synthesized using chemical route showing the promise of such an approach~\cite{Bondesgaard2019,Wu2020}. However, the relevance of the synthesis parameters or the versatility of the method was not addressed. Indeed, only easily reduced metals were considered to achieve a quinary alloy and not its subset. To achieve this goal, the target system in the present study is CoFeNiPtRu whose redox potentials and degradation temperatures for each element are given in Table~\ref{Listes}. The surfactants used are oleic acid and oleylamine (in equimolar proportion), a widely used mixture allowing to get spherical type NPs with controllable sizes. 
\begin{table*}[!ht]
\centering
\begin{tabular}{|c|c|c|c|c|}
\hline
Metallic precursors & Thermal degradation (\degree C) & Redox reaction & E$^0$/V (V)~\cite{Vanysek2005}  \\ 
\hline
\hline
Pt(acac)$_2$ & 150 - 160~\cite{Sahu2014} & Pt$^{2+} + 2e^- \rightleftharpoons$ Pt & + 1.18   \\ \hline
Ni(acac)$_2$ & 220~\cite{Chen2017} - 230~\cite{Ahrenstorf2007} & Ni$^{2+} + 2e^- \rightleftharpoons$ Ni & - 0.257  \\ \hline
Co(acac)$_2$ & 260~\cite{Kuhlman1999} & Co$^{2+} + 2e^- \rightleftharpoons$ Co & - 0.28  \\ \hline
\multirow{2}*{Ru(acac)$_3$} & \multirow{2}*{260~\cite{Mahfouz2007}} & Ru$^{3+} + 1e^- \rightleftharpoons$ Ru$^{2+}$ & + 0.249  \\ 
 & &  Ru$^{2+} + 2e^- \rightleftharpoons$ Ru & + 0.455   \\ \hline
Fe(acac)$_3$ & 180~\cite{JovicOrsini2018} & Fe$^{3+} + 3e^- \rightleftharpoons$ Fe & - 0.037  \\ 
\hline
\end{tabular}
\caption{Standard redox potentials and thermal degradation temperature of metal precursors for each element present in the targeted Fe-Co-Ni-Pt-Ru HEA NPs. Values are given at room temperature and atmospheric pressure.}
\label{Listes}
\end{table*}
Given the broad range of the quantities reported in Table~\ref{Listes}, the reduction kinetics for each cation are expected to be very different with significant consequences on the structure of the NPs. Indeed, if one element reduces before the others, it will be the first to nucleate giving rise to core-shell structures. As highlighted in case of bimetallic Ni$_{x}$Pt$_{1-x}$ NPs~\cite{MoreiraDaSilva2020}, the key point to obtain solid solution NPs is to compensate the differences in the reactivities of the metallic precursors by increasing the temperature of the synthesis. We established a linear correlation between the difference in the standard redox potentials of the cations used and the temperature required to obtain bimetallic solid solution nano-alloys~\cite{MoreiraDaSilva2020}. The synthesis is therefore carried out at a suitable temperature which ensures that all metal precursors are reduced at the same rate. To sustain this temperature, we have developed a hot injection technique where a combined solution of metal precursors and reducing agent contained in a minimum volume of solvent is rapidly injected into a much larger volume of hot solvent containing the surface stabilising agents. This process limits the cooling effect and helps to include metal atoms inside the growing NPs at the same rate throughout the synthesis to produce solid solution NPs. 

We used this approach to the synthesis of Fe-Co-Ni-Pt-Ru NPs in a FCC solid solution with controlled size. This system is very ambitious because it involves elements with very different physical properties. Indeed, it consists of transition metals crystallising individually in FCC phases with the exception of iron (bcc) and cobalt (hcp). In addition, it contains elements of different transition metal series (3$d$, 4$d$ and 5$d$) with a large disparity of atomic radii up to 20 \% but also different types of chemical bonds. Finally, the case of Ru is particularly interesting since it involves a complete reduction in two steps (see Table~\ref{Listes}), which makes the synthesis of HEA NPs via the colloidal route very challenging. In the following, it is not the aim to describe a precise synthesis of the different subsets of the CoFeNiPtRu system, but rather to show the versatility and efficiency of our method by proposing a step-by-step approach (considering some model subsets) where the influence of the experimental parameters will be discussed. As a result, we will prove how this method is particularly simple to obtain HEA NPs.
  
\section{Bimetallic alloys}

\begin{figure}[htbp!]
\begin{center}
\includegraphics[width=1.0\linewidth]{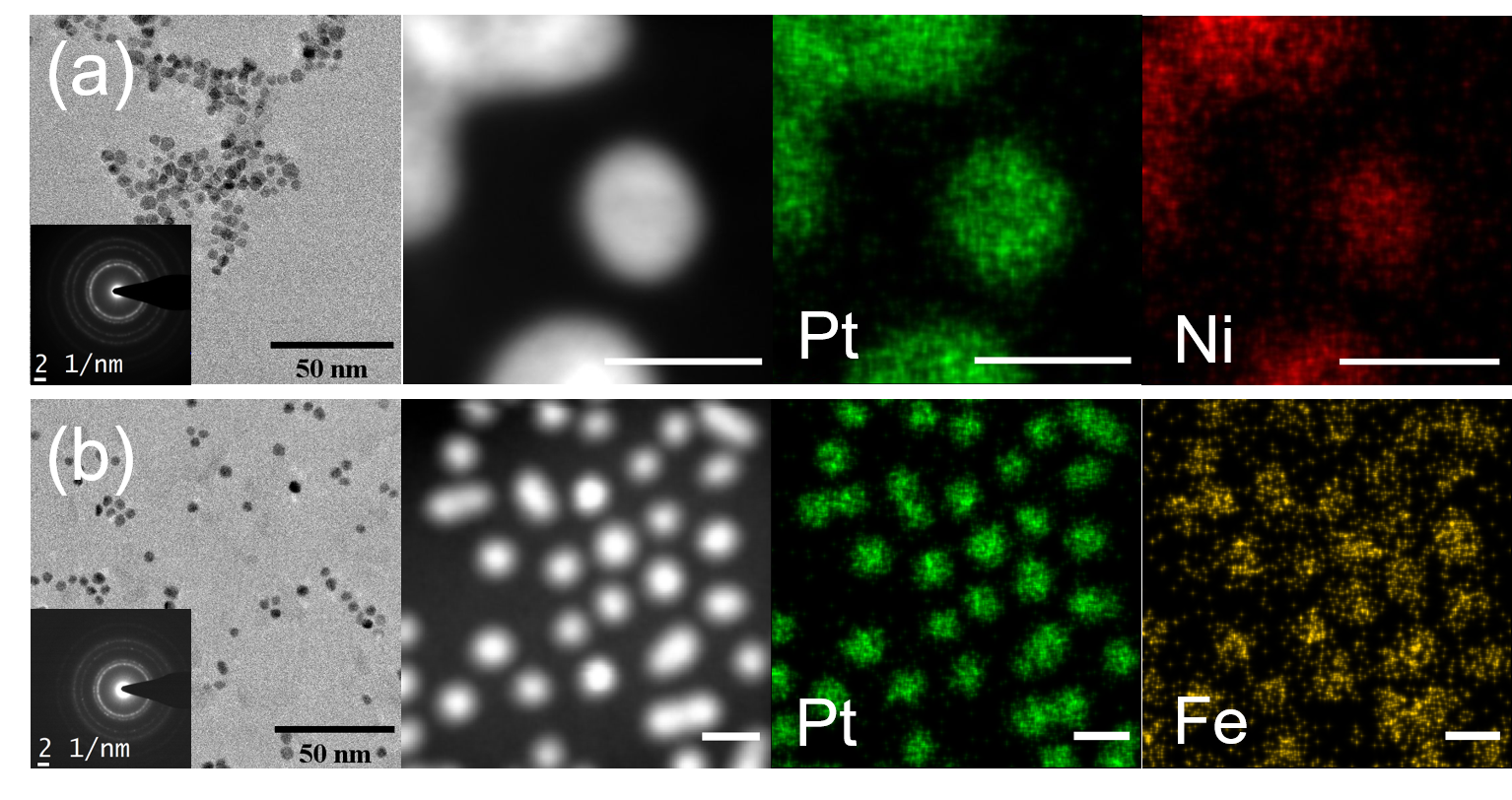}
\caption{(a) NiPt (Ni=(56$\pm$6) \%at, Pt= (44$\pm$6) \%at.) and (b) FePt (Fe=47$\pm$7 \%at., Pt= 53$\pm$7 \%at.) bimetallic NPs. (Left) Bright-field micrography of sample representative of the NP dispersion for all samples studied. As inset, electron diffraction patterns corresponding to FCC structure. (Center) HAADF-STEM image of NPs. (Right) EDX chemical mappings showing the spatial correlation between the different elements present in a set of NPs of the sample. (Scale bar = 5  nm)}
\label{fgr:bimetallique}
\end{center}
\end{figure}

In this section, two bimetallic NPs are considered, namely NiPt and FePt. For the first one, we simply recall our previous results~\cite{MoreiraDaSilva2020} where the different parameters were optimized to obtain Ni$_{x}$Pt$_{1-x}$ NPs with different compositions (Ni$_{0.75}$Pt$_{0.25}$, Ni$_{0.50}$Pt$_{0.50}$ and Ni$_{0.25}$Pt$_{0.75}$). Note that these two elements crystallise in an FCC phase in the bulk state and have lattice parameter differences of about 11 \%. As seen in Figure~\ref{fgr:bimetallique}a from transmission electron microscopy (TEM) images, NiPt NPs of equi\-atomic composition and average size of 4 nm were obtained according to the previously described hot injection technique~\cite{MoreiraDaSilva2020}. As discussed in Ref~\cite{MoreiraDaSilva2020}, a linear correlation between the difference in the standard redox potentials of the cations used ($\Delta E^0$) and the temperature required to obtain bimetallic solid solution nano-alloys is evidenced. Consequently, increasing the synthesis temperature (here 280~\degree C) results in the formation of a truly alloyed and not a core-shell structure. The homogeneous distribution of Ni and Pt atoms within the nanoparticle is confirmed from chemical mappings using energy-dispersive X-ray spectroscopy (EDX). Figure~\ref{fgr:bimetallique}a show the high-angle annular dark-field (HAADF) imaging in scanning transmission electron microscopy (STEM) mode as well as the corresponding EDX maps of Ni and Pt.  Average values of the chemical compositions extracted from the EDX spectra (Ni=(56$\pm$6) \%at, Pt= (44$\pm$7) \%at.) confirm that chemical compositions are close to the target stoechimoetry showing that all the precursors are indeed consumed during the reaction.
In order to show the robustness and reproducibility of our synthesis method, we seek to apply the same approach to Fe$_{x}$Pt$_{1-x}$ NPs. Contrary to the Ni$_{x}$Pt$_{1-x}$ cases, an additional difficulty is to obtain a fcc solid solution from iron which crystallises in the bcc phase in the bulk state. In this particular system, the chosen temperature is equal to 295~\degree C~\cite{Elkins2003} knowing that $\Delta E^0 = 1.14$~V~\cite{Vanysek2005}. A series of samples with targeted compositions Fe$_3$Pt, FePt and FePt$_3$ have been synthesized by adjusting Fe : Pt ratios (Fe$_3$Pt, and FePt$_3$ are not shown here). In case of equiatomic composition, TEM analysis and electron diffraction emphasize that NPs have an average diameter of around 4 nm and are crystallised in a disordered FCC structure (see Figure~\ref{fgr:bimetallique}b). No chemical order on the FCC lattice is observed which is expected to be the stable state in the bulk phase~\cite{Okamoto1990}. To go further EDX measurements have been performed on different areas of the TEM grids in a statistical way showing the formation of a truly alloyed where Fe and Pt atoms are randomly distributed within the NP. Interestingly, the chemical composition from the EDX analysis shows an equiatomic distribution as targeted by our synthesis.  As seen in Figure~\ref{fgr:bimetallique}b for equiatomic composition where a stoichiometry equal to Fe$=48$~\%.at, Pt$=52$~\%.at is highlighted. By considering a set of NPs, an average composition and standard deviation of composition, measured about thirty particles, equal to : Fe$=(47 \pm 7)$~\%.at, Pt$=(53 \pm 7)$~\%.at is found. 

By correlating the synthesis temperature and the redox potential of the metallic precursors, the formation of Fe$_{x}$Pt$_{1-x}$ NPs in FCC solid solution and with a given composition has been successfully achieved highlighting the versatility of our colloidal route. 

\section{Beyond bimetallic alloys}

We now focus on Co-Ni-Pt NPs of around 6 nm diameter, to make structural analysis by TEM easier. In this context, bimetallic systems synthesis was extended by dividing the amount of Ni (Ni(acac)$_2$), Pt (Pt(acac)$_2$) and Co (Co(acac)$_2$) metal precursors in three equal molar proportions.
\begin{figure}[htbp!]
\begin{center}
\includegraphics[width=1.00\linewidth]{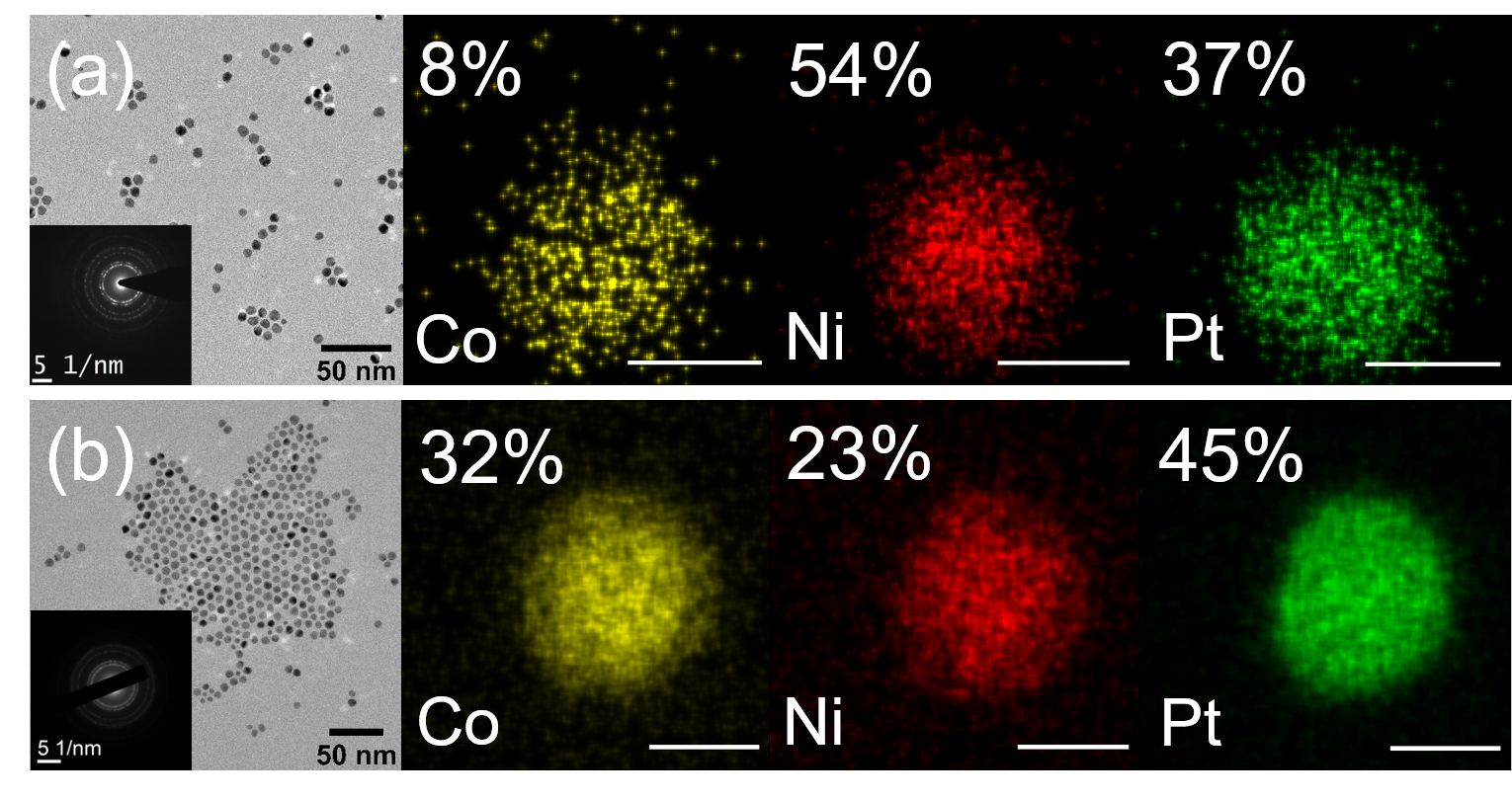}
\caption{Co-Ni-Pt NPs synthesized at (a) 280~\degree C and (b) 290~\degree C. (Left) Bright-field micrography of sample representative of the NP dispersion for all samples studied. As inset, electron diffraction patterns corresponding to FCC structure. (Right) EDX chemical mappings showing the spatial correlation between the different elements present in individual CoNiPt NPs of the sample. Concentration of each element within the NP are indicated. (Scale bar = 5 nm).}
\label{fgr:trimetallique}
\end{center}
\end{figure}
To determine the temperature required for producing Co-Ni-Pt NPs in solid solution, redox potentials of the different elements are compared. As seen in Table~\ref{Listes}, Ni$^{2+}$ and Co$^{2+}$ redox potentials are very close suggesting that the deduced temperature for Co-Ni-Pt system is therefore the same as for the Ni-Pt system, namely 280~\degree C. However, synthesis performed at this temperature leads to an inhomogeneous distribution of Co, Ni, Pt atoms within about twenty NPs (slight distribution of Ni in NP core) with a poor concentration of Co : Co$=(8\pm2)$~\%.at, Ni$=(54\pm5)$~\%.at and Pt$=(37\pm7)$~\%.at (see Figure~\ref{fgr:trimetallique}a). This failure is due to the degradation temperatures of Ni(acac)$_2$ and Co(acac)$_2$ precursors which are significantly different ($\approx 40$~\degree C as seen in Table~\ref{Listes}). This leads to disparities in degradation kinetics between the precursors during the synthesis which is an obstacle to reach nano-alloys in solid solution of equimolar composition. To overcome this bottleneck, the injection temperature of the metal precursors is increased to 290~\degree C, 10~\degree C higher than the one used for Ni-Pt system syntheses. Through this simple adjustment, a homogeneous distribution of Co, Ni and Pt atoms within about thirty NPs are obtained with an average composition close to equimolarity : Co$=(32\pm2)$~\%.at, Ni$=(23\pm8)$~\%.at and Pt$=(45\pm10)$~\%.at (Figure~\ref{fgr:trimetallique}b). Using electron diffraction, a FCC structure is clearly identified with a lattice parameter $a$ equal to $0.370 \pm 0.005$~nm as expected by the Vegard law. Moreover, no ordered structures have been observed. \\
\begin{figure}[htbp!]
\begin{center}
\includegraphics[width=1.0\linewidth]{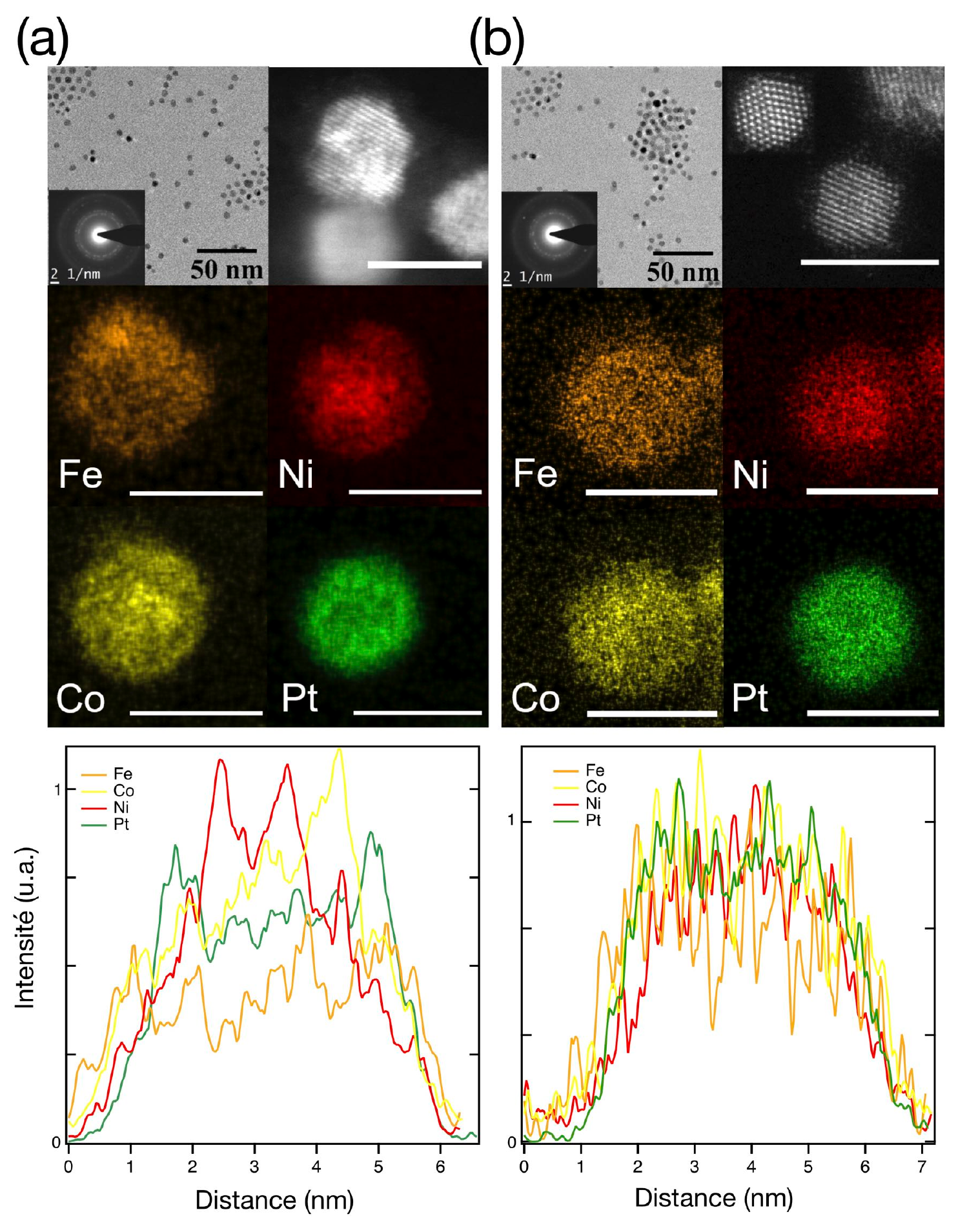}
\caption{Fe-Co-Ni-Pt NPs. To keep the temperature constant at  297~\degree C, the quantity of precursors is reduced during the synthesis from (a) Fe=14$\pm$1 \%at., Co= 22$\pm$2 \%at., Ni= 27$\pm$2 \%at., Pt= 37$\pm$4 \%at. (around twenty particles). (b) Fe=12$\pm$2 \%at., Co= 19$\pm$4 \%at., Ni= 26$\pm$5 \%at., Pt= 43$\pm$8 \%at. (around thirty particles). (Top) Bright-field micrography of sample representative of the NP dispersion for all samples studied. As inset, electron diffraction patterns corresponding to FCC structure. HAADF-STEM image (high-resolution mode) of NPs showing an inhomogeneous distribution within the NP for (a) and perfectly crystalline homogeneous structures as well as twinned NP.  (Middle) EDX chemical mappings showing the spatial correlation between the different elements present in individual FeCoNiPt NPs of the sample. (Bottom) Intensity profiles of all elements along the radius of the NP.  (Scale bar = 5 nm)}
\label{fgr:quadrametallique_1}
\end{center}
\end{figure}
After this successful synthesis of the ternary system, the same procedure including an additional metal element is performed. Our choice lies with the Co-Fe-Ni-Pt alloy, where iron is added with the supplementary difficulty of incorporating a bcc-type element to obtain a fcc solid solution. As the Fe$^{III}$(acac)$_3$ precursor degrades and reduces according to the same mechanisms as the precursors used so far, the synthesis temperature (here T~$= 297$~\degree C) is simply adapted according to the $\Delta E^0$. As seen in Figure~\ref{fgr:quadrametallique_1}a, the resulting NPs display the same characteristics as the previous ones. Indeed, we obtain NPs of monodisperse diameter $d=(4.7 \pm 1.3 )$~nm and of FCC type. However, the chemical composition analyses show a non-homogeneous dispersion of the different elements within the NPs as seen in STEM-HAADF image whose contrast is related to $Z$, the atomic number difference between elements. The chemical mappings using EDX mode coupled to the intensity profiles along the radius of the NP confirm this trend by revealing a homogeneous distribution of Fe and Co and a core and surface enriched in Ni and Pt, respectively. Again, the role of temperature was found to be very important.  Indeed, we found necessary to increase the volume of the injected solution to keep the mixture of the four metallic precursors soluble. This resulted in a strong temperature drop after the injection, explaining the previously described features for the nanoparticles. Under these conditions, a sufficiently high temperature is maintained throughout the synthesis process. In order to overcome this difficulty, the quantity of precursors was reduced and therefore the volume too so that so that the temperature is maintained sufficiently high. As seen in Figure~\ref{fgr:quadrametallique_1}b, all metallic elements are now homogeneously distributed within the NP in a disordered FCC structure. HAADF-STEM images exhibits a homogeneous contrast consistent with a random distribution of elements within the NP. Moreover, they also reveal the presence of perfectly crystalline nanoparticles but also many twinned ones. From the chemical mapping analysis and the intensity profiles along the NPs, we can clearly state that no segregation is reported. Consequently, the colloidal route is perfectly adapted to produce solid solution alloyed FCC tetrametallic NPs with controlled size.

\begin{figure*}[htbp!]
\begin{center}
\includegraphics[width=1.00\linewidth]{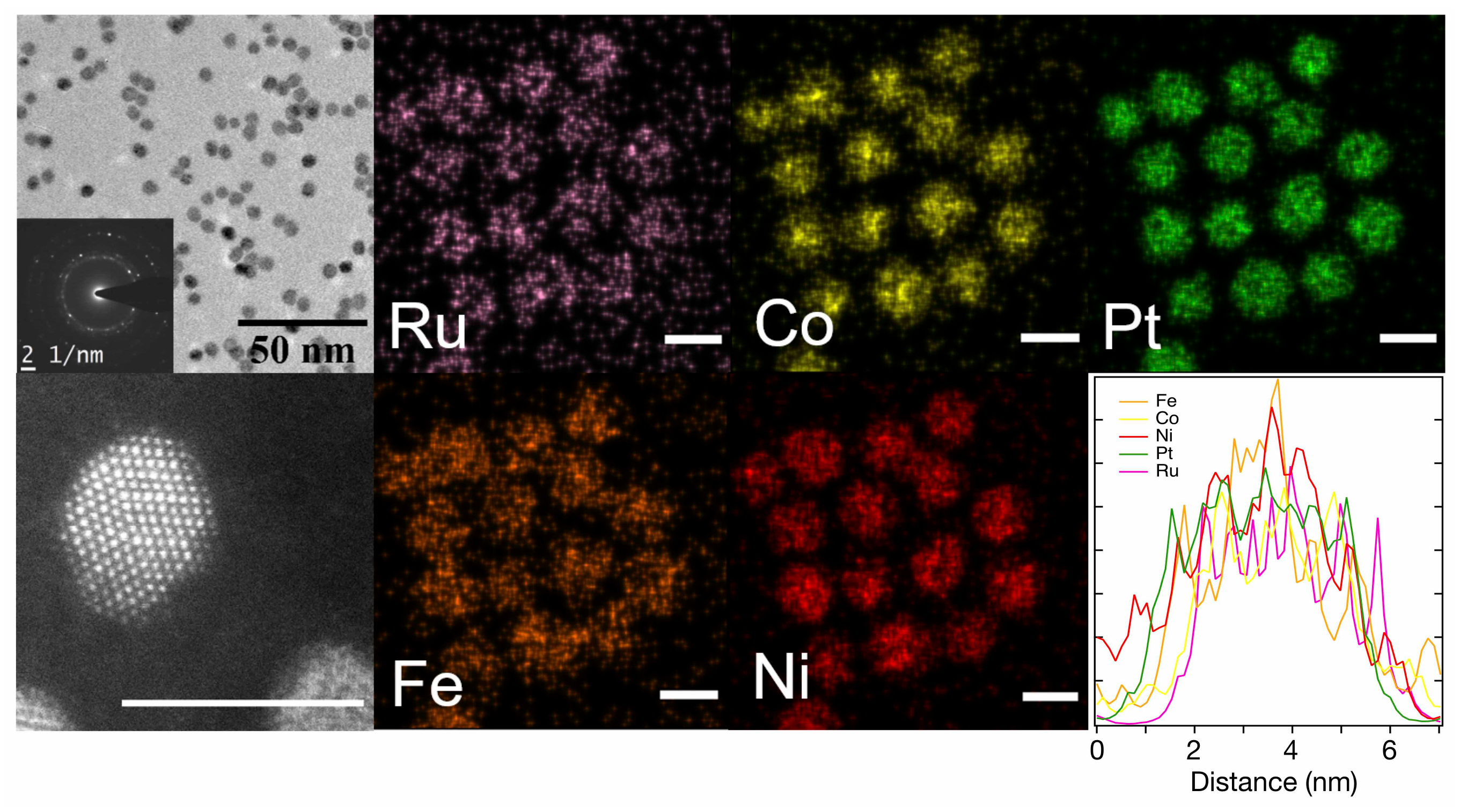}
\caption{Fe-Co-Ni-Pt-Ru NPs of composition : Fe=12$\pm$2 \%at., Co= 22$\pm$2 \%at., Ni= 23$\pm$2 \%at., Pt= 37$\pm$2 \%at., Ru= 6$\pm$3 \%at. (measure on about twenty particles.) (Left) Bright-field micrography of sample representative of the NP dispersion for all samples studied. As inset, electron diffraction patterns corresponding to FCC structure. HAADF-STEM image (high-resolution mode) of NP. (Middle) EDX chemical mappings showing the spatial correlation between the different elements present in individual FeCoNiPtRu NPs of the sample. (Left) Intensity profiles of all elements along the radius of the NP.  (Scale bar = 5 nm)}
\label{fgr:pentametallique}
\end{center}
\end{figure*}
To further emphasize the versatility of our method, a ruthenium precursor (Ru(acac)$_3$) is added to the previous system. As its acetylacetonate complex, Ru is in its +III oxidation state. Unlike the other species, this requires a transition to an intermediate +II state (well known for its kinetical inertness) before reaching its zero oxidation (see Table~\ref{Listes}). Such a particular feature may \textit{a priori} be a real difficulty in optimising the synthesis conditions to ensure a homogeneous distribution of the Ru within the NP and not a possible segregation. This difficulty was overcome thanks to the experience acquired on different systems studied earlier. According to comparison of the redox potentials involved in the Fe-Co-Ni-Pt-Ru system, the injection temperature was set at 290~\degree C. As seen in Figure~\ref{fgr:pentametallique}, our colloidal approach leads to the formation of pentametallic NPs  in a straightforward manner. Fe-Co-Ni-Pt-Ru NPs have the same characteristics as the previous ones: a FCC structure where no order is identified. STEM-HAADF image reveal the formation of perfectly crystallised NPs. Thanks to EDX analysis in mapping mode (Figure~\ref{fgr:pentametallique}), the presence of all metal atoms within NPs is highlighted. No phase segregation is observed which satisfies the definition of an HEA.

\section{Conclusion}
In summary, we have presented a simple, fast and successful method for producing HEA NPs. Despite the presence of many parameters that are necessarily correlated, we have shown that it is possible to optimise the synthesis conditions very easily in order to obtain NPs with an FCC structure and homogeneously distributed elements. We have proven the efficiency of the colloidal route for the Fe-Co-Ni-Pt-Ru system, quite a challenging one considering the structural and chemical differences of all the constituting elements. There is no doubt that our method can be extended to other intrinsically simpler systems, even if this implies some adjustments of the experimental parameters, \textit{i.e.} by comparing the reducibility of metal cations/synthesis temperature. Typically, the control of the chemical composition within the NP is a relevant point but it is beyond the scope of the paper. Indeed, our present work clearly demonstrates the viability of the colloidal route to make HEAs in a versatile way based on the correlation between the difference in redox potentials and temperature. This should be the guideline to adjust the experimental conditions to obtain NPs in FCC solid solution of controlled size and chemical composition. Interestingly, its simplicity of use, implementation and also its relatively low cost make the colloidal route perfectly suited to meet the requirements of industrial applications of NPs~\cite{Stark2015}. This work constitutes therefore a major step in the  mastery of HEA NPs synthesis which is still a challenge. The colloidal route opens up new prospects for the development of new families of materials with as yet unsuspected properties.
% Experimental section

% Acknowledgements
\acknowledgments
The authors thank ANR GiANT (N\degree ANR-18-CE09-0014-04) and ARF "Nano" of ONERA for the financing of this work. HRSTEM-EDX study was carried out within the MATMECA consortium, supported by the ANR-10-EQPX-37 contract and has benefited from the facilities of the Laboratory MSSMat (UMR CNRS 8579), CentraleSup\'elec. H.A. would like to warmly thank Damien Alloyeau, Jaysen Nelayah and Christian Ricolleau for the fruitful discussions but also for their unfailing support in initiating this project.

\appendix

\section{Nanoparticules synthesis}
The procedure used to synthesize multimetallic nanoalloys is an extension of the method described elsewhere~\cite{MoreiraDaSilva2020}. To prevent contaminations, all laboratory glasswares were washed by \textit{aqua regia} (nitric acid 1 : 3 hydrochloric acid) during (at least) five hours and rinsed with large amounts of distilled water, dichloro\-methane (DCM) and acetone. Synthesis  were carried out under argon blanket. Commercial reagents were used without purification. Typically, to obtain NPs with 5.5 to 6.5~nm in diameter, a first solution of 35~mL of benzyl ether (BE) (Sigma Aldrich, 99~\%), 0.2~mL of oleylamine (OAm) (Acros Organics, 80~-~90~\%) and 0.2~mL of oleic acid (OAc) (Acros Organics, 80~-~90~\%) was loaded in a 100 mL round-bottom flask containing a PTFE coated magnetic stirring bar. The mixture was purged by three vacuum/argon cycles and heated during 10~min at 100~\degree C, to remove any water trace and prevent particle oxidation. The temperature was then raised and kept at the synthesis temperature. By the same time, a second solution of 0.33~mmol of acetyl or acetylacetonate metallic precursors divided in equimolar proportion between the various constituents (here Fe(III) acetylacetonate (Sigma Aldrich, 97~\%), Co(II) acetylacetonate (Sigma Aldrich, 97~\%), Ni(II) acetylacetonate (Sigma Aldrich, 95~\%), Pt(II) acetylacetonate (Sigma Aldrich, $\geq$~99.98~\%), Ru(III) acetylacetonate (Sigma Aldrich, 97~\%)) and 3.1~mmol of 1,2-hexadecanediol (TCI~98~\%) in 1.3~mL of BE, 0.2~mL of OAc and 0.2~mL of OAm was prepared under vigorous stirring. This second solution was then quickly added to the first heated one in the round-bottom flask. The resulting solution became instantly black, proof of NPs nucleation. The suspension was kept during 8~min before cooling it under an argon blanket down to 40~\degree C. Then, 20~mL of ethanol (EtOH) for 10~mL of the mixture were then added, and the resulting suspension was centrifuged at 10~000~rpm for 20~min. The supernatant was separated and the precipitate was washed again with EtOH and centrifuged. The precipitate was redispersed in dichloro\-methane for further characterizations.

\section{TEM analysis}

As-synthesized NPs were characterized by the Transmission Electron Microscopy (TEM) technique. A drop of colloidal suspension was deposited and dried on a copper TEM grid with a suspended thin carbon film. Size distribution (on population of 500~particles in 5 distinct zones on TEM grid and counted with Image J software) and electron diffraction (ED) were measured using a TFS-CM 20 TEM (200~kV). High-Resolution Transmission Electron Microscopy (HRTEM), High-Angle Annular Dark-Field imaging (STEM-HAADF) and STEM/chemical mapping were performed using a Titan G2 Cs-corrected FEI TEM operating at 200~kV on individual particles. (Nickel, Iron, Cobalt) and (Platinum, Ruthenium) signals are obtained from their K$_{\alpha}$ and L$_{\alpha}$ line intensities, respectively. The results of the chemical compositions obtained on the isolated particles are verified by EDX spectra obtained on large areas containing about 100 particles.

% References
% Use the following code if you wish to generate your bibliography with BibTeX;
% replace the string "MSP-template" below with the name(s) of
% the BibTeX data base(s) you want to use.
% The resulting bibliography-output (the content of the .bbl file)
% must be pasted back into this file before submission.
% Please also include your BibTeX data base file(s) in your submission
% so that we can re-run BibTeX if necessary.
%

%

\end{document}